\documentclass[12pt]{iopart}

\usepackage{lipsum}
\usepackage{graphicx}
\usepackage[dvipsnames]{xcolor}

\begin{document}

\title{Exclusion model of mRNA translation with collision-induced ribosome drop-off}

\author{Johannes Keisers$^{1}$ and Joachim
  Krug$^2$\footnote{Corresponding author (jkrug@uni-koeln.de).}}
\address{$^1$ Centre de Biologie Structurale (CBS), Universit\'e de Montpellier, CNRS, INSERM, 29 rue de Navacelles 34090, Montpellier, France}
\address{$^2$ Institute for Biological Physics, University of Cologne,
  Z\"ulpicher Strasse 77, 50937 K\"oln, Germany}
\begin{abstract}
The translation of messenger RNA transcripts to proteins is commonly modeled as a one-dimensional totally asymmetric exclusion process with extended particles. Here we focus on the effects of premature termination of translation through the irreversible detachment of ribosomes. We consider a model where the detachment is induced by the unsuccessful attempt to move to an occupied site. The model is exactly solvable in a simplified geometry consisting of the translation initiation region followed by a single slow site representing a translation bottleneck. In agreement with recent experimental and computational studies 
we find a non-monotonic dependence of the ribosome current on the initiation rate, but only if the leading particle in a colliding pair detaches. Simulations show that the effect persists for larger lattices and extended bottlenecks. 
In the homogeneous system the ribosome density decays asymptotically as the inverse square root of the distance to the 
initiation site. 
\end{abstract}
%\keywords{Exclusion models, translation, collision-based drop-off}

\submitto{\jpa}
\maketitle

\section{Introduction}
Many natural processes occur far from equilibrium, but a general framework for describing non-equilibrium phenomena is still lacking. A promising approach to understanding the relation between microscopic and macroscopic behavior in non-equilibrium systems is to use simple models, such as driven diffusive systems \cite{Kriecherbauer2010,Chou2011,Schadschneider2011}.

A paradigmatic driven diffusive system is the one-dimensional totally asymmetric simple exclusion process or TASEP, 
where particles move uni-directionally and stochastically along the lattice subject to an exclusion interaction, which implies that multiple
particles are not allowed to occupy the same site. Even before the TASEP was formally introduced to the probabilistic literature by Spitzer \cite{Spitzer1970}, it had been used by MacDonald et al. for modeling the polymerization of RNA and polypeptides from their respective macromolecular templates~\cite{MacDonald1968}. Specifically, when describing the 
translation of messenger RNA (mRNA) to proteins, the particles represent ribosomes moving along an RNA transcript. One is thus lead to consider TASEP's with extended particles
covering $\ell > 1$ sites on a finite lattice with open boundaries 
\cite{MacDonald1968,Shaw2003,Zia2011}. The rates of injection ($\alpha$) and extraction $(\beta$) 
of particles at the boundaries of the 
lattice correspond to the initiation and termination rates of the translation process, and the particle current determines the rate of protein synthesis. In general, the 
elongation 
rates at which ribosomes advance along
the lattice depend on the underlying RNA sequence, and the hopping rates of the TASEP are therefore inhomogeneous \cite{Zur2016,Erdmann2020,Szavits2020,Josupeit2021}. 

For point particles ($\ell = 1$) and homogeneous hopping rates the TASEP is exactly solvable~\cite{Derrida1992,Derrida1993,Schuetz1993,Krug2016}. The solution displays various boundary-induced phase transitions in the $(\alpha,\beta)$-plane \cite{Krug1991} which persist also for extended particles ($\ell > 1$) \cite{Shaw2003}. In particular, when translation is limited by initiation, the ribosome current increases with the
initiation rate $\alpha$ up to a critical value $\alpha^\ast$ and saturates for larger $\alpha$. In the regime
$\alpha > \alpha^\ast$ the current is limited by the collisions between the particles.

Translation initiation is widely recognized as the rate-limiting step in protein production, and the majority of translation processes lead to the production of functional proteins~\cite{li2014}. However, elongation stalls can occur during translation, necessitating ribosome rescue mechanisms and subsequent degradation of non-functional or defective proteins~\cite{keiler2015}. 
The \textit{processivity} of a ribosome refers to its ability to efficiently and accurately complete the entire translation process without encountering errors or disruptions. 
In bacteria, it is estimated that processivity errors occur at a rate of $4 \times 10^{-4}$ per codon~\cite{Jorgensen1990,Kurland1992,Sin2016}.
There are many mechanisms through which the premature termination of translation can be
induced, and they are employed to ensure the fidelity of protein synthesis and maintain quality control over the translated proteins \cite{Ikeuchi2019a}. Following \cite{Sin2016},
we refer to these processes collectively as \textit{ribosome drop-off}. 

Within the TASEP framework, the limited processivity of ribosomes violates the conservation of particles in the bulk of the lattice, and places the process in the class of exclusion models with Langmuir kinetics originally introduced for describing the traffic of motor proteins on cellular filaments \cite{Popkov2003,Parmeggiani2004}. In the latter context particles may attach to
as well as detach from the lattice, whereas in the present setting of mRNA translation, detachment is irreversible. Previous work on TASEP's with Langmuir kinetics \cite{Popkov2003,Parmeggiani2004,Pierobon2006,Bonnin2017} has generally assumed that detachment and (if applicable) attachment occurs randomly, without any coupling to the local environment of the particle (but see \cite{Rakos2003}). Here we focus instead on a scenario
where the detachment is induced by particle collisions.

Experimental studies using ribosome profiling \cite{Ingolia2014} have established that ribosome collisions are widespread in translation \cite{diament2018}, and they may elicit a wide range of cellular responses when they occur \cite{Meydan2021}. For example, experiments with bacteria that were starved for specific amino acids suggest that the formation of queues behind stalled ribosomes may be avoided by the detachment of ribosomes from the mRNA \cite{Subramaniam2014}. 
Importantly, there is mounting evidence in eukaryotes \cite{Simms2017,Juszkiewicz2018,Ikeuchi2019} and bacteria \cite{Ferrin2017,Leedom2022} 
that ribosome collisions can be detected by the cell as signatures of processivity errors
and subsequently trigger ribosome drop-off, possibly in conjuction with cleavage of the 
mRNA. The latter effect will not be considered here.

Our work is specifically motivated by a recent investigation by Park and Subramaniam \cite{Park2019}, who modified the initiation region of the mRNA sequence of the \textit{PGK1} gene in yeast in such a way that the translation initiation rate could be varied systematically. In addition, a stall sequence was inserted at the end of the gene, which serves as a translation bottleneck and allows to control and localize the number of ribosome collisions. Remarkably, the experiment revealed a decrease in protein expression when the initiation rate was large enough, a novel and counterintuitive phenomenon. Using a detailed computational model of the translation process, Park and Subramaniam showed that the observed decrease of the protein production could be reproduced by assuming a collision-dependent drop-off process, but only if the leading (rather than the trailing) ribosome in a colliding pair drops off. Additionally, their work revealed that collisions between ribosomes not only affect translation dynamics but also decrease the stability of the mRNA.

The goal of this article is to introduce an
exclusion model of translation with collision-induced ribosome drop-off, which provides an analytic explanation for the surprising qualitative difference between drop-off processes affecting the leading or the trailing ribosome that was observed in~\cite{Park2019}.
Under the assumption that the bulk of the collisions occur at the bottleneck, we consider a minimal model for the dynamics at a bottleneck comprising a single site, which turns out to be exactly solvable. 
We subsequently use simulations to study additional effects arising
from extended bottlenecks, as well as from the dynamics in the regions of the mRNA sequence flanking the 
bottleneck. The homogeneous version of our model is related to previously studied asymmetric reaction-diffusion models 
with a localized input of particles \cite{Cheng1989,Lebowitz1996,Hinrichsen1997,Ayyer2010}, and we exploit this connection to infer the asymptotic behaviour of the particle density in large systems.

\section{Model}
\label{Sec:Model}

\subsection{Definition}
\begin{figure}
    \centering
    \includegraphics[width=\textwidth]{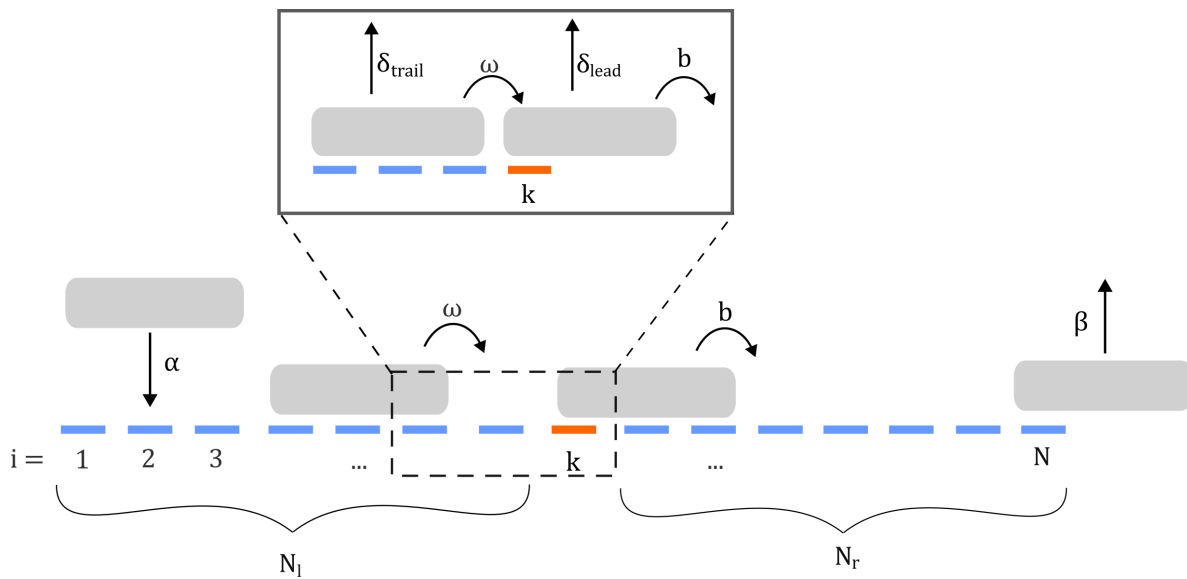}
    \caption{\textbf{Schematic depiction of the model.} The one-dimensional lattice is separated into three regions: A bottleneck of size $n$ ($n = 1$ in the figure), and two flanking regions of sizes $N_l$ and $N_r$ in front and behind the bottleneck. The total number of sites is $N = N_l + N_r + n$. A particle of size $\ell$ can enter the lattice at the initiation rate $\alpha$ if the first $\ell$ sites are not occupied. The hopping rates, $\omega$, are homogeneous except for the bottleneck rate $b < \omega$, which is applied at positions $i=k, k+1, \dots, k+n-1$. Particles at site $N$ leave the system at the termination rate $\beta$. The cutout in the black box illustrates the collision-induced drop-off process, as well as the minimal model considered in Section~\ref{Sec:minimalmodel}. If a hopping particle is blocked in the forward direction, the trailing and the leading particle drops off with probability $\delta_\mathrm{trail}$ and $\delta_\mathrm{lead}$, respectively. 
    }
    \label{fig:fig1}
\end{figure}

Ribosome collisions occur preferentially when the ribosome encounters a translational bottleneck. Therefore, the model used in this work is an inhomogeneous TASEP with open boundaries, particle size $\ell \geq 1$ and a bottleneck region, as illustrated in Figure~\ref{fig:fig1}. The dynamics are specified by the initiation rate $\alpha$, the baseline hopping rate $\omega$, the bottleneck rate $b < \omega$, and the termination rate $\beta$. Following standard terminology in 
the field \cite{Zur2016,Erdmann2020,Bonnin2017}, we refer to the rate at which the particles jump to the next site as the elongation rate. 
The jump rate of a particle is determined by the position of its leftmost site \cite{Zia2011,Shaw2004}.

In addition, following \cite{Park2019}, we introduce a collision-induced drop-off process where the leading or trailing particle of a colliding pair is removed from the system after a failed elongation attempt. The probabilities for these (generally independent) processes are denoted by $\delta_\mathrm{trail}$ and $\delta_\mathrm{lead}$, respectively. In our model, we assume that drop-off events are solely triggered by particle collisions. Therefore, a drop-off event occurs only when the elongation attempt of the trailing particle is unsuccessful due to the presence of another particle. The probabilities $\delta_\mathrm{trail}$ and $\delta_\mathrm{lead}$ represent the likelihood that a failed elongation attempt leads to a particle drop-off event. It is important to note that, in this scenario, the drop-off rate cannot surpass the elongation rate.

The piling up of particles in front of the bottleneck leads to an increased number of collisions, and correspondingly to a partial localisation of the drop-off events in this region. 
In order to better understand the consequences of the two modes of collision induced drop-off in front of the bottleneck, we consider a minimal model, shown in the cutout box in Figure~\ref{fig:fig1}, where the entire lattice consists of the initiation region and a bottleneck of size $n = 1$. Because the size of the initiation region is equal to the particle size $\ell$, particles can only collide if the lattice is fully occupied. In the minimal model, the initiation rate $\alpha$ represents the rate at which particles arrive at the bottleneck, and the bottleneck rate 
$b$ effectively plays the role of the termination rate $\beta$. 

The analytic solution of the minimal model in Section~\ref{Sec:minimalmodel} is complemented by simulations
using the standard Gillespie algorithm~\cite{GILLESPIE1976}. Unless otherwise stated, each simulation was run for $10^7$ time steps to reach the steady state, and subsequently another $10^7$ time steps were performed for collecting data. Our focus on the stationary behaviour is justified by the observation that, in models of this type, the steady state is typically reached before mRNA degradation becomes relevant \cite{Szavits2020}.  

\subsection{Exact solution of the minimal model}
\label{Sec:minimalmodel}
As illustrated in Figure~\ref{fig:fig1}, the minimal model is defined on a lattice of size $N = \ell +1$.
The system can be in $N+2$ different states: The empty lattice (state $e$), the full lattice occupied by two particles (state $f$), and states $i = 1, \dots, N$ where a single particle occupies site $i$. Denoting the corresponding state probabilities by $P_e, P_f$ and $\{P_i\}_{i=1,\dots,N}$, the master equation governing the system reads as follows:
\numparts
\begin{eqnarray}
    &\dot{P_e} = b P_{N} - \alpha P_e, \label{eq:1a}\\
    &\dot{P_1} = \alpha P_e  + (b +\omega \delta_\mathrm{lead})P_f - \omega P_1, \label{eq:1b} \\
    &\dot{P_i} = \omega P_{i-1} - \omega P_{i} \qquad \mathrm{for} \qquad i =  2, ..., \ell , \label{eq:1c}\\
    &\dot{P_N}= \omega P_{N-1}+ \omega \delta_\mathrm{trail} P_f - (\alpha + b) P_{N}, \label{eq:1d} \\
    &\dot{P_f} = \alpha P_{N} - (b + \omega \delta_\mathrm{trail} + \omega \delta_\mathrm{lead})P_f. \label{eq:1e}
\end{eqnarray}
\endnumparts
The steady state is obtained by setting the right hand sides of these equations to zero. As a consequence
of equation (\ref{eq:1c}) the stationary single particle probabilities $\bar{P}_i$ are all equal for $i=1, \dots \ell$. 
Making use of this simplification and the normalization condition $\bar{P}_e + \ell \bar{P}_1 + \bar{P}_N + \bar{P}_f = 1$, the stationary distribution can be written down straightforwardly, and the 
stationary particle current is obtained from the relation
\begin{equation}
    J = b (\bar{P}_f+\bar{P}_{N}).
\end{equation}

We now specialise to the situation where only one of the two drop-off processes is active, and 
set either $\delta_\mathrm{lead} = 0$ or $\delta_\mathrm{trail} = 0.$
When the drop-off affects the leading particle the stationary current is given by
% \numparts
\begin{equation}
    J_\mathrm{lead} = b \frac{1 + \frac{\alpha}{\omega \delta_\mathrm{lead} + b}}{1 + \frac{b}{\alpha}+ \frac{\alpha}{\omega \delta_\mathrm{lead} + b} + \frac{\ell}{\omega}(\alpha + b)}, \label{eq:Jlead}
    \end{equation}
and when the trailing particle drops off one obtains
    \begin{equation}
    J_\mathrm{trail} = b \frac{1 + \frac{\alpha}{\omega \delta_\mathrm{trail} + b}}{1 + \frac{b}{\alpha} + \frac{\alpha}{\omega \delta_\mathrm{trail}+ b} + \frac{\ell b}{\omega} \left(1 + \frac{\alpha}{b + \omega \delta_\mathrm{trail}}\right)}. \label{eq:Jtrail}
\end{equation}
% \endnumparts
Park and Subramaniam \cite{Park2019} refer to the two modes of drop-off described by equations (\ref{eq:Jlead}) and 
(\ref{eq:Jtrail}) as \textit{collision-stimulated abortive termination} (CSAT) and 
\textit{collide and abortive termination} (CAT), respectively.

\begin{figure}[h]
    \centering
    \includegraphics[width=0.5\textwidth]{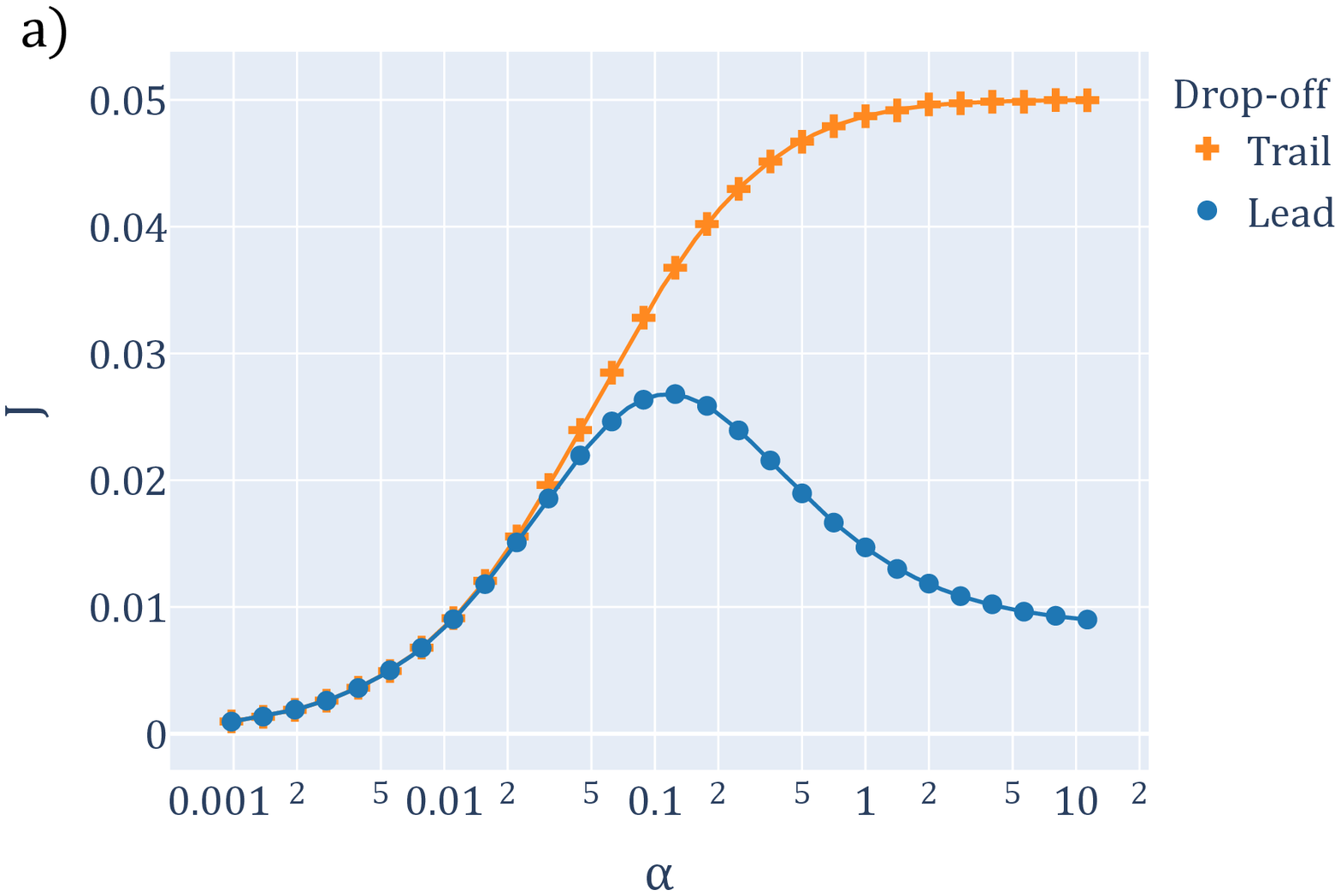}\includegraphics[width=0.5\textwidth]{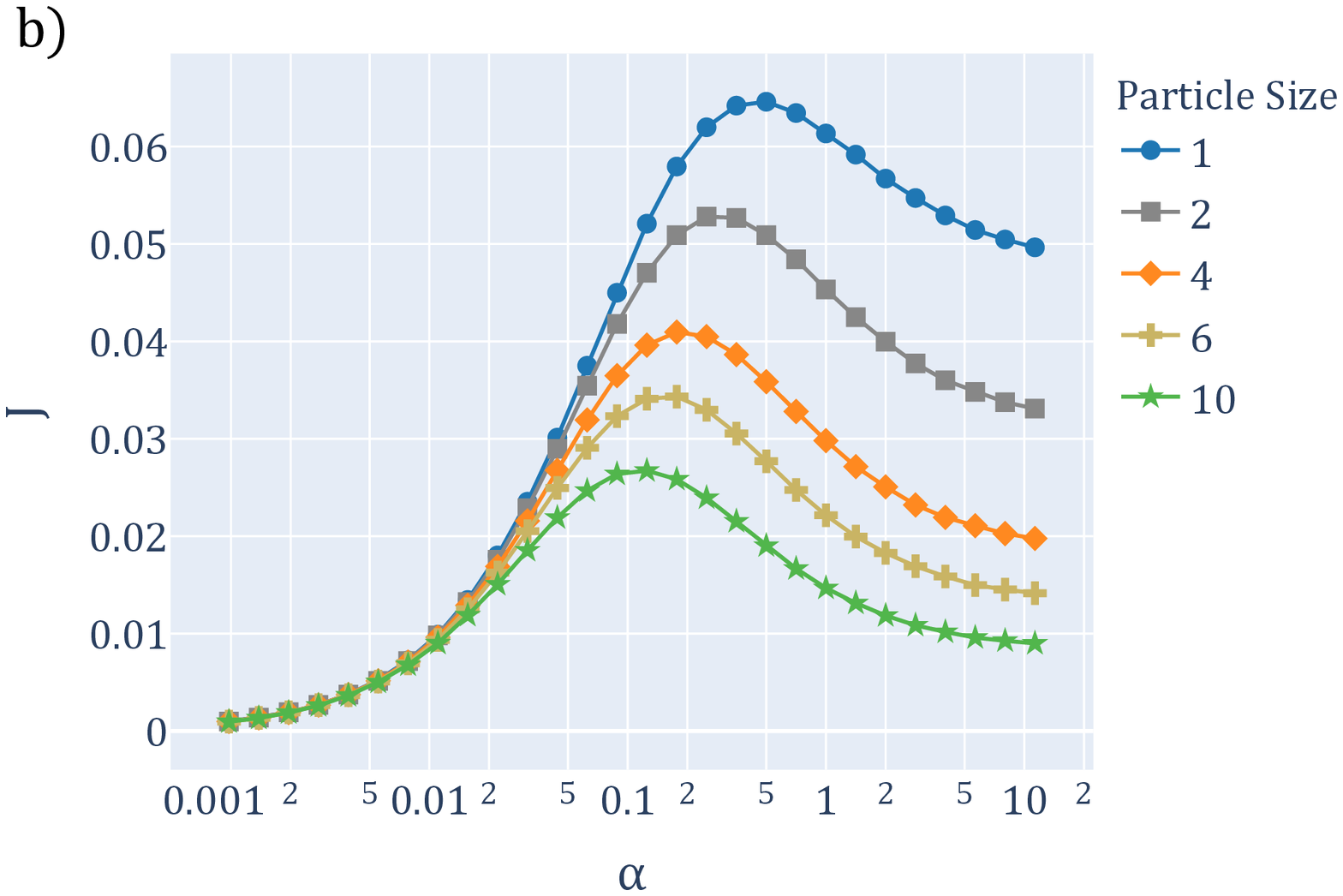}
    \caption{\textbf{Stationary current in the minimal model.} The figure compares the analytic expressions (\ref{eq:Jlead}) and (\ref{eq:Jtrail}) (full lines) to simulation results (symbols). In both panels the initiation rate $\alpha$ is varied on a logarithmic scale, and the elongation and bottleneck rates are set to $\omega = 1$, $b=0.1$. a) Comparison of the current with trailing and leading particle drop-off for particles of size $\ell = 10$. In both cases the probability $\delta$ of the operative drop-off mechanism is set to unity. b) Impact of the particle size $\ell$ on the current $J_\mathrm{lead}$ with maximal drop-off probability ($\delta_\mathrm{lead} = 1$). Increasing the particle size lowers the overall current, but the degree of non-monotonicity increases.}
    \label{fig:fig2}
\end{figure}

For $\ell=1$ and $\delta=0$, equations (\ref{eq:Jlead}) and (\ref{eq:Jtrail}) reduce to the stationary current of the standard TASEP on two sites ($N=2$) which was previously derived in \cite{Derrida1993}. The difference between the two expressions becomes apparent in the limit of large initiation rate, $\alpha \rightarrow \infty$, where they reduce to
\numparts
\begin{eqnarray}
    J^{\infty}_\mathrm{lead} = \frac{b}{1+ \ell \left(\frac{b}{\omega} + \delta_\mathrm{lead} \right)},\label{eq:4a}\\
    J^{\infty}_\mathrm{trail} = \frac{b}{1+ \ell \frac{b}{\omega} }.\label{eq:4b}
\end{eqnarray}
\endnumparts
Thus the drop-off process reduces the current for large $\alpha$ only if it affects the leading particle. 
As a consequence, 
as seen in Figure~\ref{fig:fig2}, $J_\mathrm{lead}$ is non-monotonic with respect to $\alpha$, whereas 
$J_\mathrm{trail}$ increases monotonically. In the next section we discuss the underlying mechanism
in detail.
 
\section{Non-monotonic dependence of ribosome current on initiation rate}

\subsection{Stationary current in the minimal model}

As shown in the previous section, the difference between the two drop-off mechanisms has a major impact on the stationary current. The current for both mechanisms is obviously the same for very low initiation rates. However, upon increasing the initiation rate, particles collide on the lattice creating a collision-induced drop-off current. 

In general, the current in an inhomogeneous exclusion process is set by the slowest rate in the system~\cite{Zia2011,Josupeit2021}. Thus, when the initiation rate exceeds the bottleneck rate, the current is limited by the flux through the bottleneck. This also means that if the initiation rate is approximately equal to the bottleneck rate, particles start to collide at the bottleneck. The difference between the two mechanisms is that in the case of the leading particle drop-off, the collisions induce a drop-off current at the bottleneck site, which competes with the current leaving the system. On the other hand, the trailing particle drop-off creates a drop-off flux before the bottleneck. In the regime $\alpha > b$, the bottleneck is constantly occupied and determines the current of the system. Contrary to leading particle drop-off, which reduces the occupancy of the bottleneck, trailing particle drop-off does not. Therefore, $J_\mathrm{trail}$ does not change after the initiation rate is high enough to constantly fill the bottleneck site.

Equation (\ref{eq:Jlead}) shows that
the degree of the non-monotonic behaviour of $J_\mathrm{lead}$ depends on the bottleneck rate, the drop-off probability, and the particle size. 
The position of the maximum is at $\alpha \approx b$ and marks a trade-off point between the number of collisions and the initiation rate. The current only shows a significant decrease for particle size $\ell > 1$, as seen in equation (\ref{eq:4a}) and Figure~\ref{fig:fig2}b). Conceptually, when the leading particle drops off, the frontmost particle is moved back by one particle length, and $\ell$ elongation steps have to be performed to restore the previous situation. Thus, the particle size represents the current reduction per drop-off event. 

We conclude that the minimal complexity required for the current to display a non-monotonic dependence on the initiation rate is a system consisting of an initiation region and a single slow site, with a collision-induced drop-off mechanism where the leading particle drops off. Moreover, the degree of non-monotonicity depends strongly on the particle size. In the remainder of this section we focus on the drop-off mechanism affecting the leading particle, and set $\delta_\mathrm{trail} = 0$ throughout.

\begin{figure}
    \centering
    \includegraphics[width=0.5\textwidth]{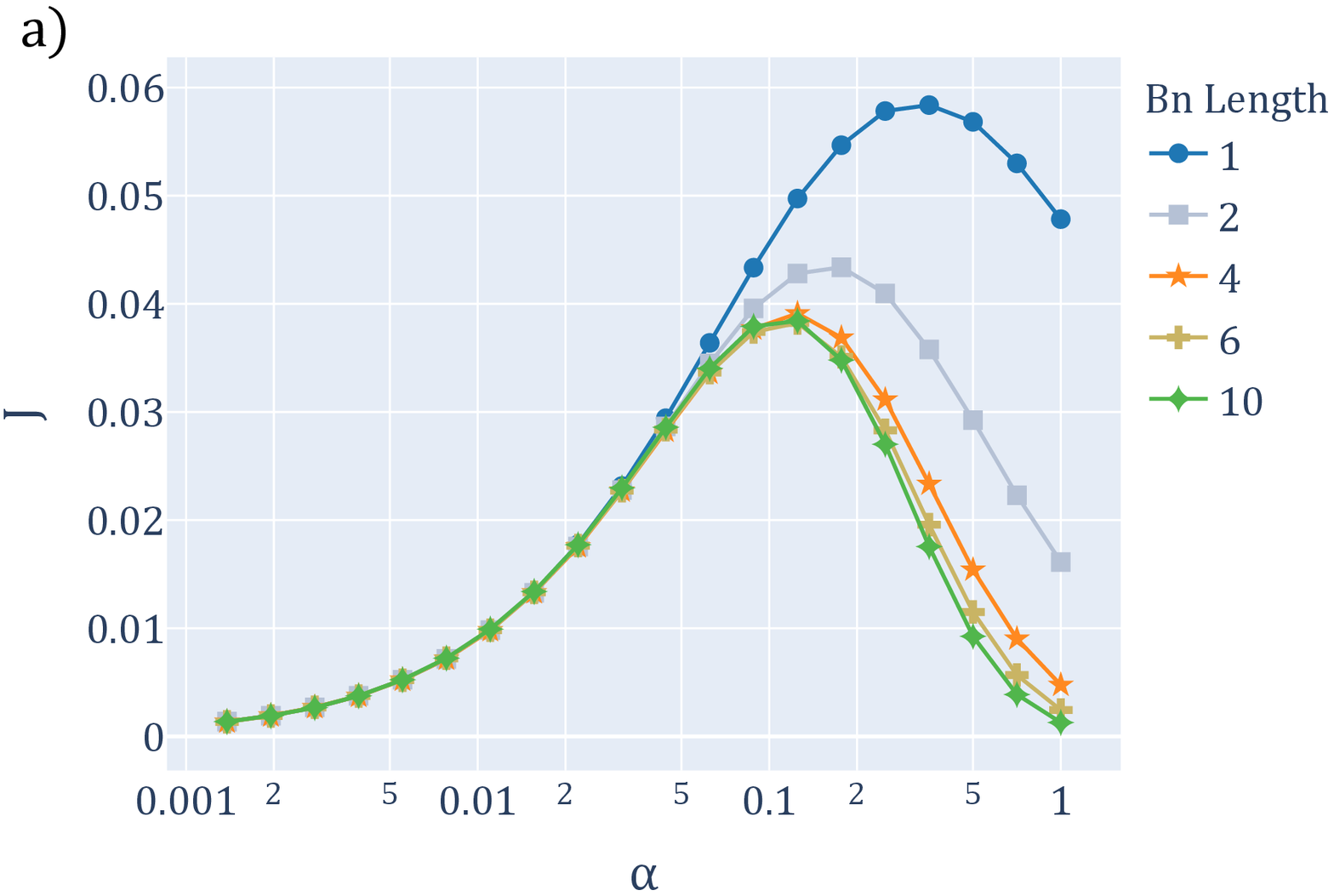}\includegraphics[width=0.5\textwidth]{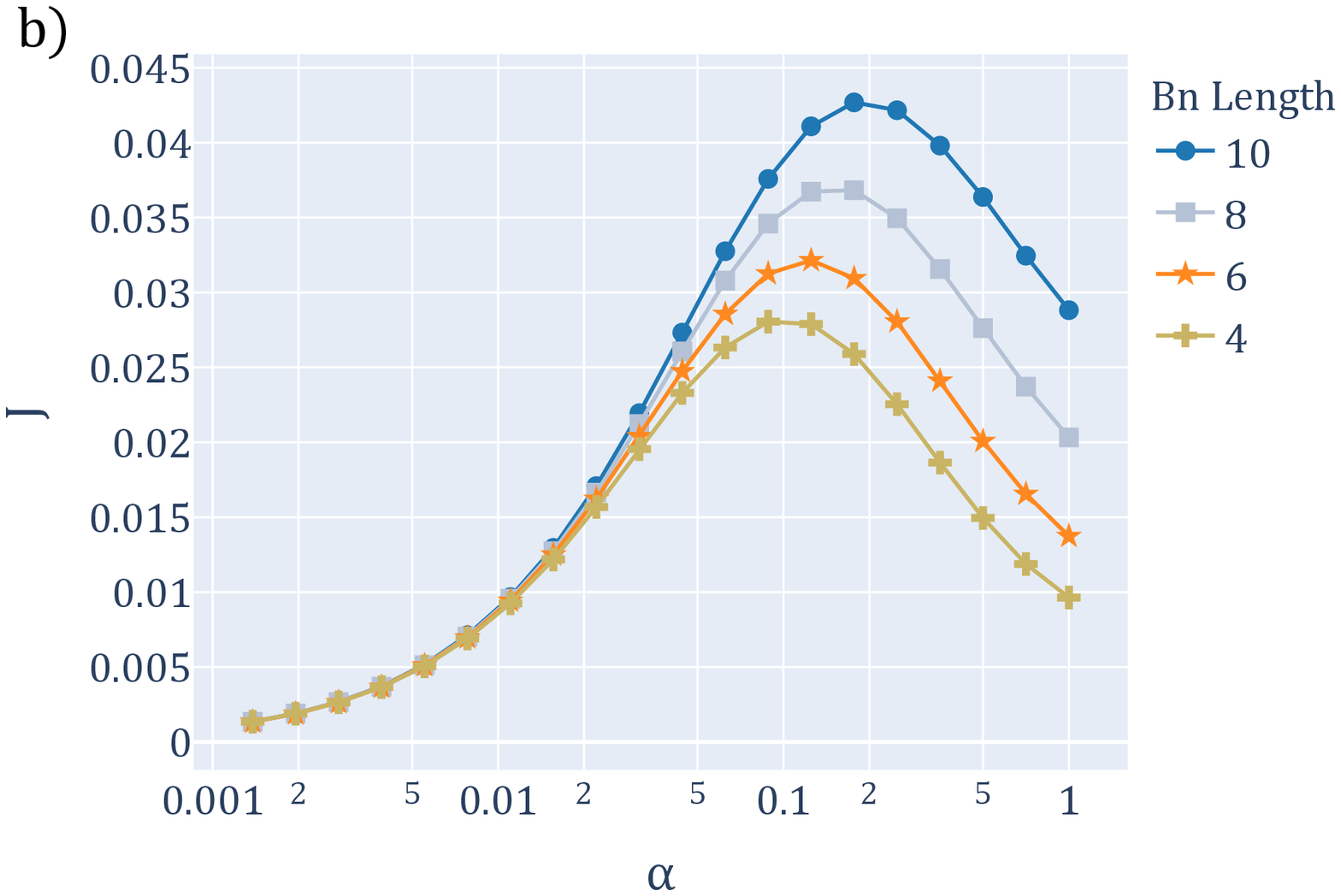}
    \caption{\textbf{Effect of the bottleneck length on the stationary current.} The figure shows simulation results for the stationary current as a function of the initiation rate for a system consisting of an initiation region of length $\ell=10$ and bottlenecks of different lengths $n$. The average time it takes a particle to cross the bottleneck is kept constant by multiplying the elongation rate in the bottleneck region with its length, $b_n = n b_1$ with $b_1 = 0.1$.  a) For strong, short bottlenecks, the current decreases and the degree of non-monotonicity increases with increasing bottleneck length. Here 
    $\omega = 10 $ and $n= 1, 2, 4, 6, 10$. The effect of increasing the bottleneck length is reduced for longer bottlenecks, i.e, the change in the current is largest between $n=1$ and 2. b) For longer and/or weaker bottlenecks the trend is reversed, and the current increases with increasing length. 
    For the case $\omega =1$ shown in the figure, the reversal occurs at $n=3$, and the bottleneck effectively disappears at $n_c = \omega/b_1 = 10$. In both panels $\delta_\mathrm{lead} = 1$.}
    \label{fig:fig3}
\end{figure}

\subsection{Effects of the bottleneck length}
In the experiments of Park and Subramaniam~\cite{Park2019}, the stalling sequences consisted of 10 codons,
and in their simulations the authors observed that the suppression of the protein production at large 
initiation rate becomes more pronounced with increasing bottleneck size. As a first generalisation of the 
minimal model, we therefore consider a system consisting of the initiation region and a bottleneck of size
$n > 1$. The total lattice size is thus $N = \ell + n$.

In order to separate the effects of bottleneck strength and bottleneck size, we follow the approach
of \cite{Park2019} and rescale the per-site elongation rate in the bottleneck by $n$. 
In general, in the continuous time setting used here, the waiting time a particle spends on a given lattice site $i$ is exponentially distributed with parameter $\omega_i$, where $\omega_i$ is the hopping rate at $i$. Thus, the average time a particle stays on a lattice site $i$ is equal to $1/\omega_i$. Therefore, in order to keep the bottleneck strength constant, the elongation rate $b_n$ per site in a bottleneck region of size $n$ has to be multiplied by $n$, 
\begin{equation}
    \label{Eq:bn}
b_n = n b_1,
\end{equation}
such that the total average time $n/b_n = 1/b_1$ for crossing the bottleneck does not change.
Importantly,
the variance of the crossing time decreases with increasing $n$ \cite{Moffitt2014}. 
As seen in Figure~\ref{fig:fig3}, for strong bottlenecks the non-monotonic behaviour of the current becomes more pronounced with increasing $n$, but this trend is reversed for long bottlenecks. This is to be expected based on the following observation: For a given value of $b_1$, there is a critical bottleneck length 
$n_c = \omega/b_1$ where the elongation rate per site becomes equal to the background rate $\omega$, and hence the bottleneck effectively disappears. The applicability of (\ref{Eq:bn})
is therefore limited to $n < n_c$.

\subsection{Simulations of the full-scale model}

\begin{figure}
    \centering
    \includegraphics[width=\textwidth]{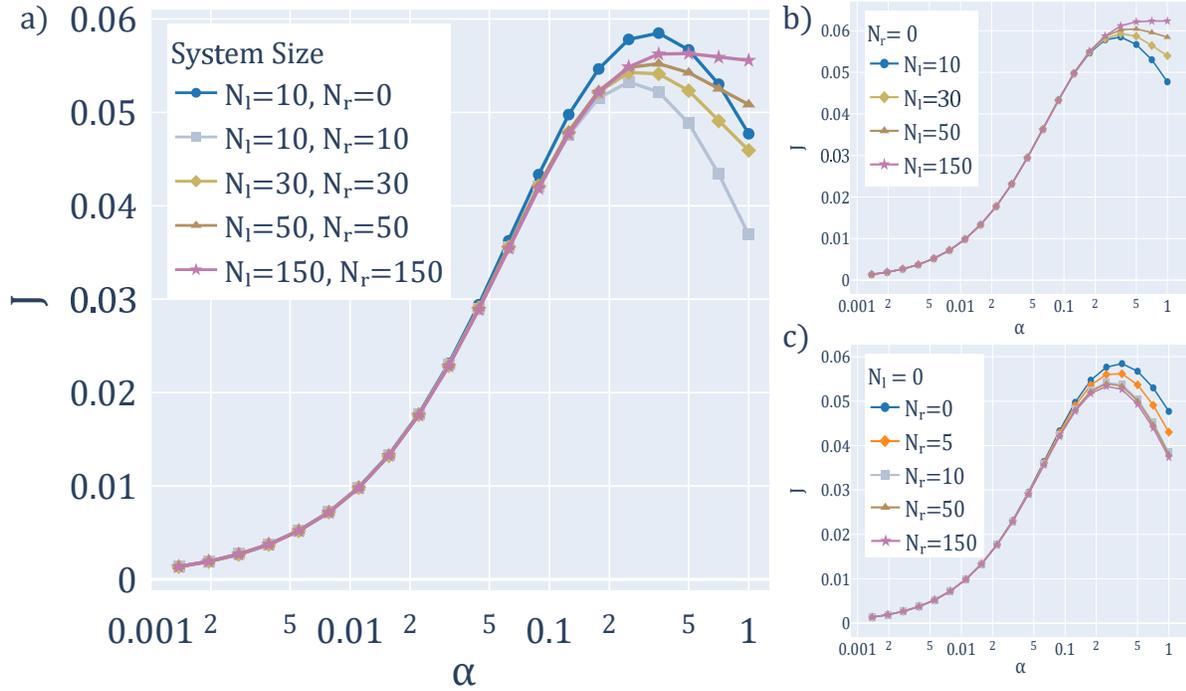}
    \caption{\textbf{Stationary current for the full-scale model.} The figure shows simulation results for the stationary current obtained for lattices of varying size comprising a single-site bottleneck ($n=1$) of strength $b=0.1$. The other parameters are $\omega = 10$, $\beta = 1$, $\delta_\mathrm{lead} = 1$ and $\ell = 10$.  a) Starting from the minimal model ($N_l = \ell = 10$, $N_r = 0$), sites are added symmetrically on both sides of the bottleneck. 
    b) Increasing $N_l$ while keeping $N_r = 0$ leads to a decrease in the non-monotonicity of the current, because collision-induced drop-off events reduce the arrival rate of particles at the bottleneck, effectively decreasing the particle drop-off. c) Increasing the system size by increasing $N_r$ while keeping $N_\ell = \ell = 10$ leads to a decrease of the current. 
    Combining the effects shown in b) and c) leads to the overall behaviour seen in a). }
    \label{fig:fig4}
\end{figure}

Although the minimal model provides insights into the mechanism behind the observed decrease of the current at large initiation rates, the lattice comprising only an initiation region and a bottleneck does not represent a typical mRNA molecule with several hundred codons. In this section we therefore use simulations to explore the full-scale version of our model, where regions of varying lengths $N_l$, $N_r$ are added on both sides of the bottleneck. The results displayed in Figure~\ref{fig:fig4}a) show that the current for the minimal and full-scale models coincides up to a certain initiation rate. The deviation from the full scale model for larger values of $\alpha$ can be explained by the different consequences of adding sites in front and behind the bottleneck. 

Figure~\ref{fig:fig4}b) shows the effect of adding sites in front of the bottleneck. By increasing $N_l$ while keeping $N_r = 0$ the non-monotonicity of the current is reduced. The degree of the non-monotonic behaviour depends on the number of drop-off events at the bottleneck. In the regime $\alpha > b$, the current is set by the number of particles flowing through the bottleneck. By adding sites in front of the bottleneck, the system allows for drop-off events before the particles reach the bottleneck. This reduces the arrival rate of particles at the bottleneck, i.e. the initiation rate in our minimal model is effectively reduced. This results in less collisions at the bottleneck and therefore less particle drop-off at the bottleneck, which, in the regime where the flux is dictated by the amount of particles flowing through the bottleneck, eventually results in the disappearance of the non-monotonicity as seen in Figure~\ref{fig:fig4}b). If the system does not display the non-monotonic behavior, the drop-off flux at the bottleneck is negligible and the system is limited by the arrival rate of the particles at the bottleneck. In terms of the minimal model, introduced in Section~\ref{Sec:minimalmodel}, the initiation rate is lower than the bottleneck rate. 

On the other hand, increasing $N_r$ leads to a reduction of the current, as depicted in figure~\ref{fig:fig4}c). The decrease is only noticeably for $1 \leq N_r \leq 10$ and increasing $N_r$ further does not impact the current significantly. Due to the low density caused by the bottleneck, the number of collisions after the bottleneck is low. But because the particles can extend beyond the bottleneck, particles on the sites $k-\ell \leq i < k$ are able to cause particles on the sites $k \leq i \leq k+\ell$ to drop off. Combining both of these effects, the deviations between the minimal model and the full scale model can be explained.

\begin{figure}
    \centering
    \includegraphics[width=\textwidth]{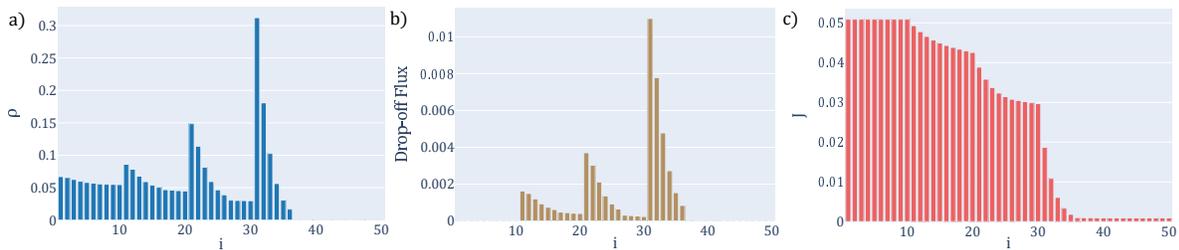}
    \caption{\textbf{Stationary profiles of density, drop-off flux and current.} The figure shows the spatial profile of a) the particle density, b) the drop-off flux, and c) the particle current, for a system of $N=50$ sites and parameters $\alpha = 0.125, \, \beta = 1, \, \omega = 1, \, \delta_\mathrm{lead} = 1$. The system contains a bottleneck of length $n=6$ at positions $i=31-36$ with a per-site elongation rate $b_n = 6 \times b_1 = 0.06$.  
    The effect of the drop-off is strongest inside the bottleneck region, where all three observables display a negative gradient. Because the particles pile up in front of the bottleneck, the sites located at multiples of $\ell$ before the bottleneck show a similar behaviour as the bottleneck sites. 
    }
    \label{fig:fig5}
\end{figure}

Both effects can also be seen qualitatively in Figure~\ref{fig:fig5}, where profiles of particle density, drop-off flux and current are depicted for a system of $N = 50$ sites and a bottleneck of length $n=6$. All three observables show a negative gradient inside the bottleneck. Moreover, the effect of the bottleneck propagates 
to the sites in front of the bottleneck in a discrete pattern reflecting the particle size.

\section{Universal large scale behaviour}

\begin{figure}[b]
    \centering
    \includegraphics{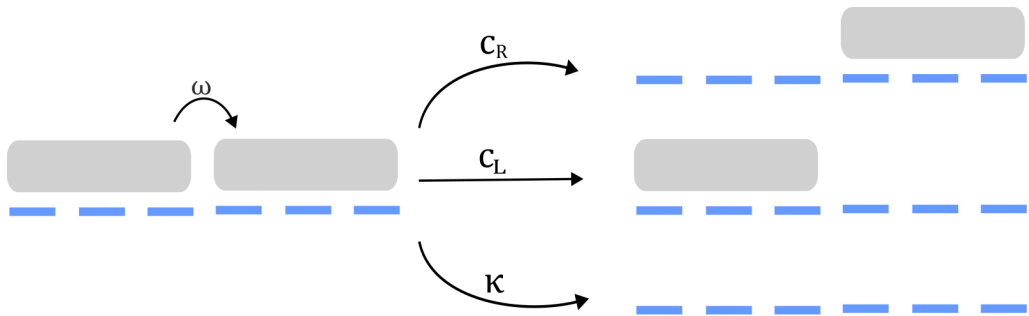}
    \caption{\textbf{Effective coalescence and annihilation rates.} The figure illustrates
    the interpretation of collision-induced particle drop-off as an effective coalescence or annihilation process with rates given by (\ref{eq:rates}). With probability 
    $(1-\delta_\mathrm{trail})(1-\delta_\mathrm{lead})$ the initial configuration remains unchanged.}
    \label{fig:fig6}
\end{figure}

The preceding section has shown that the non-monotonic dependence of the particle current on the initiation rate is reduced by adding lattice sites in front of the bottleneck and eventually disappears when $N_l$ becomes large. This raises the question about the behaviour of the model, and the difference between the two modes of collision-induced drop-off, 
in the limit $N \to \infty$. In the present subsection we address this question in the light of previous work on asymmetric reaction-diffusion models with particle coalescence or annihilation. 
For simplicity, we consider the homogeneous system without a bottleneck and with a constant bulk elongation rate $\omega$.

Our starting point is the observation that, from the viewpoint of the particles that remain on the lattice, a collision-induced 
drop-off event effectively results in the coalescence (if one particle in the pair drops off) or annihilation (if both particles
drop off) of the two neighbouring particles (Figure~\ref{fig:fig6}). The alternative drop-off of the trailing or leading particle corresponds to the 
two biased coagulation processes considered in \cite{Hinrichsen1997}. 
The effective rates of coalescence to the right or left ($c_R$, $c_L$) and the effective rate of annihilation ($\kappa$) of the 
collision-induced drop-off model are thus given by 
\begin{equation}
    \label{eq:rates}
    c_R = \omega\delta_\mathrm{trail} (1 - \delta_\mathrm{lead}), \; c_L = \omega\delta_\mathrm{lead} (1 - \delta_\mathrm{trail}), \; \kappa = \omega\delta_\mathrm{trail} \delta_{\mathrm{lead}}.
\end{equation}
In particular, when only one of the drop-off probabilities is nonzero there are no annihilation events, whereas for $\delta_\mathrm{trail} = \delta_\mathrm{lead} = 1$, coalescence events are absent and all collisions result in the removal of both particles. 

Exact solutions are available for coalescing point particles ($\ell = 1$) with rates $c_R = \omega, c_L = \kappa = 0$ 
\cite{Hinrichsen1997} and for annihilating point particles with $\kappa = \omega, c_L = c_R = 0$ \cite{Lebowitz1996,Ayyer2010}. 
These solutions predict that the stationary particle density $\rho_i$ at site $i$, $1 \ll i \ll N$, decays asymptotically as
\begin{equation}
    \label{eq:density}
    \rho_i \approx \frac{1}{\sqrt{\pi i}} \;\; \mathrm{(coalescence)} 
    \;\;\;\;\;\;\;\;  \rho_i  \approx \frac{1}{2 \sqrt{\pi i}} \;\; \mathrm{(annihilation)}.
\end{equation}
Moreover, based on simulation results, Hinrichsen et al. 
conjecture that the leading order decay law for the particle density is universal with respect to changes in the coalescence
rates, and they provide a similarity transformation that maps coalescence models to models where both coalescence and annihilation
events can occur \cite{Hinrichsen1997}. 

\begin{figure}[h]
    \centering
    \includegraphics[width=\textwidth]{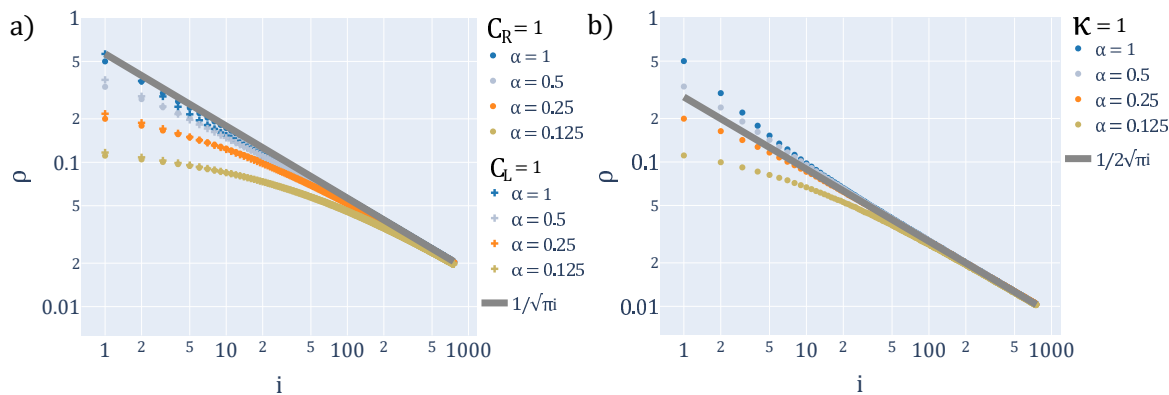}
    \caption{\textbf{Stationary density profile in a large system with homogeneous rates.} The figure shows 
    simulation results for a homogeneous system of $N=750$ sites and parameters $\ell = \beta = \omega = 1$. a) Stationary density profiles for either trailing (circles) or leading (crosses) particle drop-off, and different values of the initiation rate $\alpha$. The line shows the prediction for the coalescence model from Eq (\ref{eq:density}). Asymptotically the stationary density is independent of the initiation rate and the drop-off mechanism. b) Stationary profiles for the model where both the leading and trailing particle drops off, for different $\alpha$. The line shows the prediction for the annihilation model from Eq (\ref{eq:density}). In both panels $\alpha$ decreases from top to bottom.
    }
    \label{fig:fig7}
\end{figure}
\begin{figure}[h]
    \centering
    \includegraphics[width=0.33\textwidth]{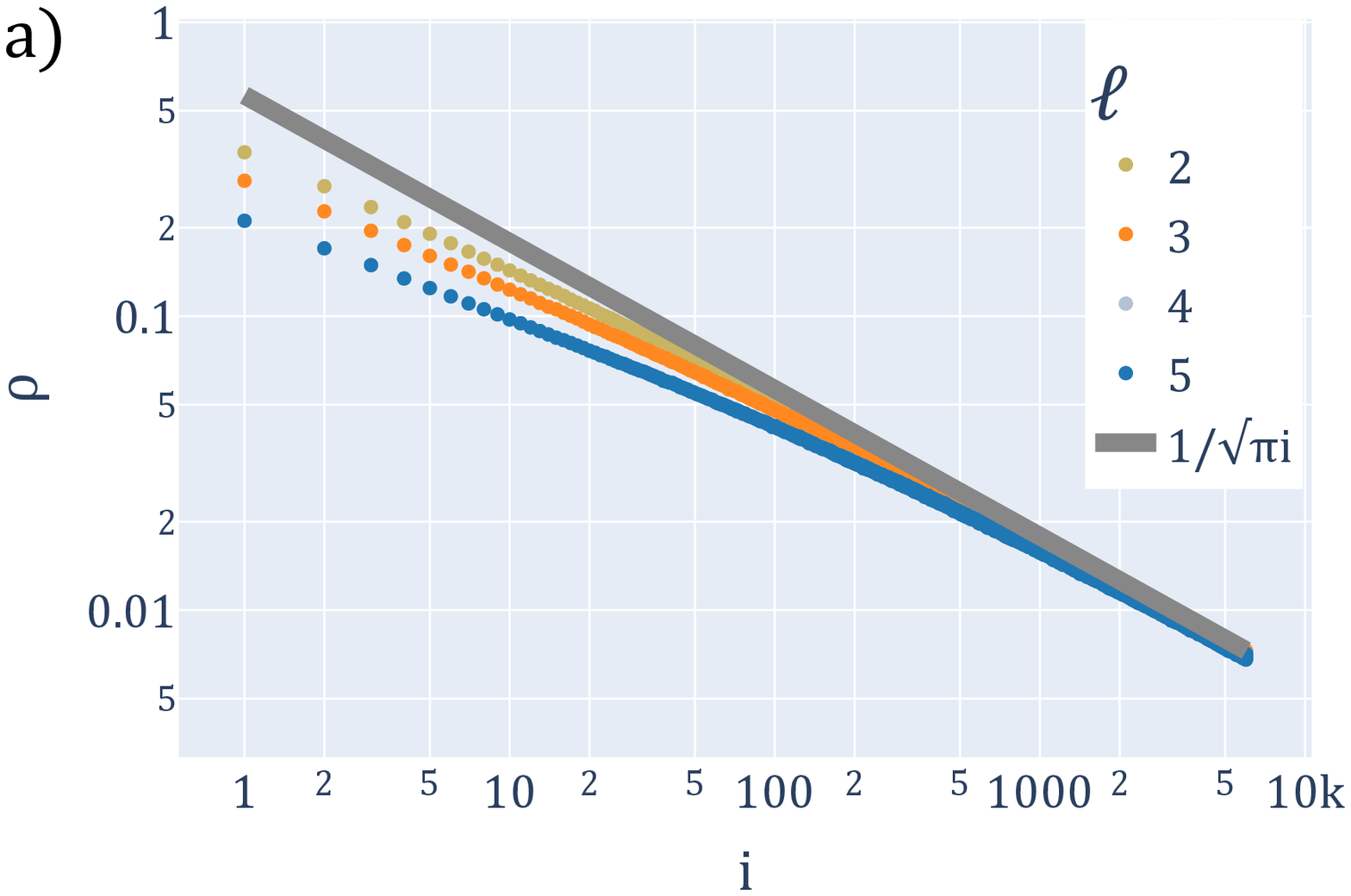}\includegraphics[width=0.33\textwidth]{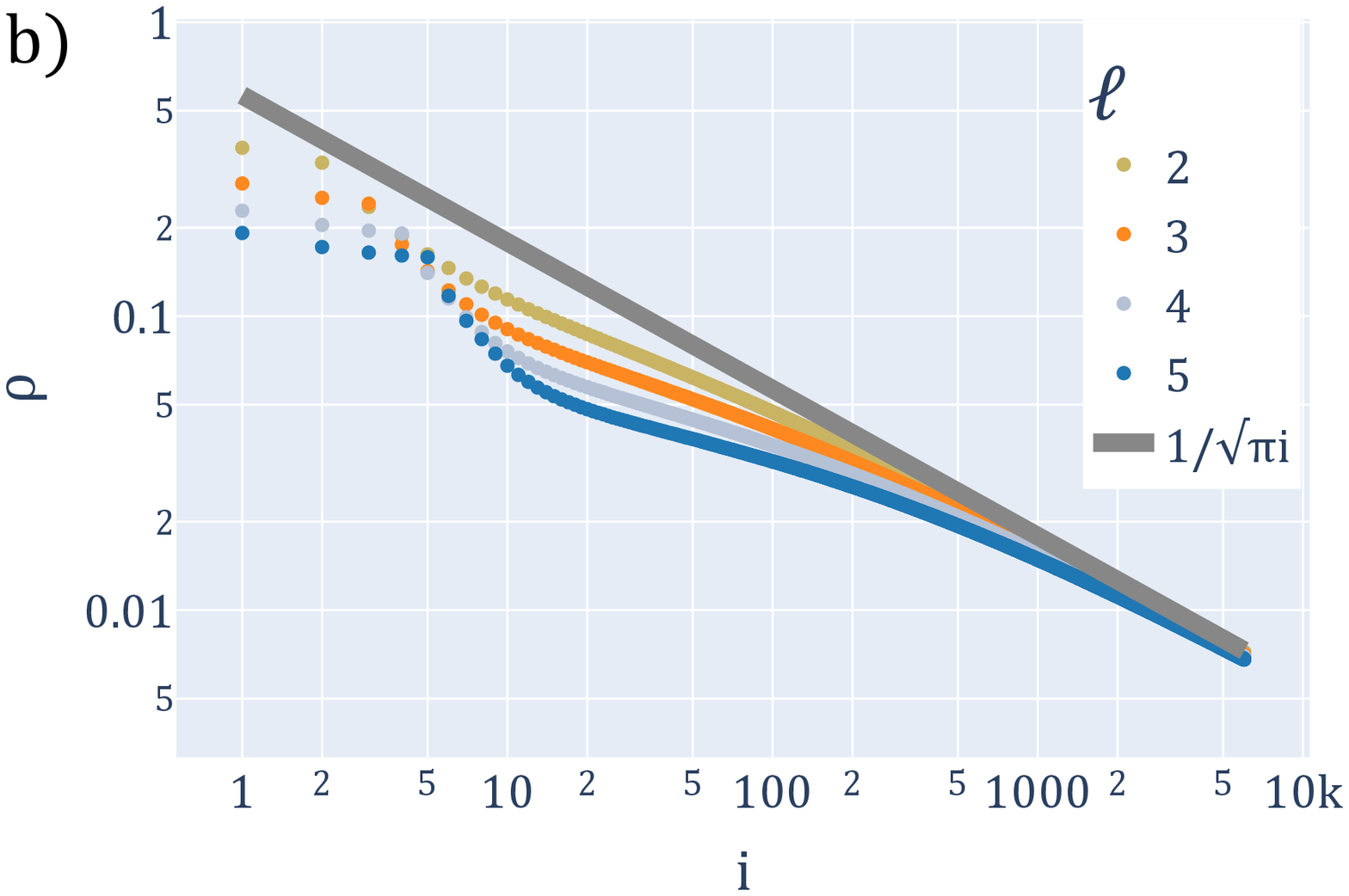}\includegraphics[width=0.33\textwidth]{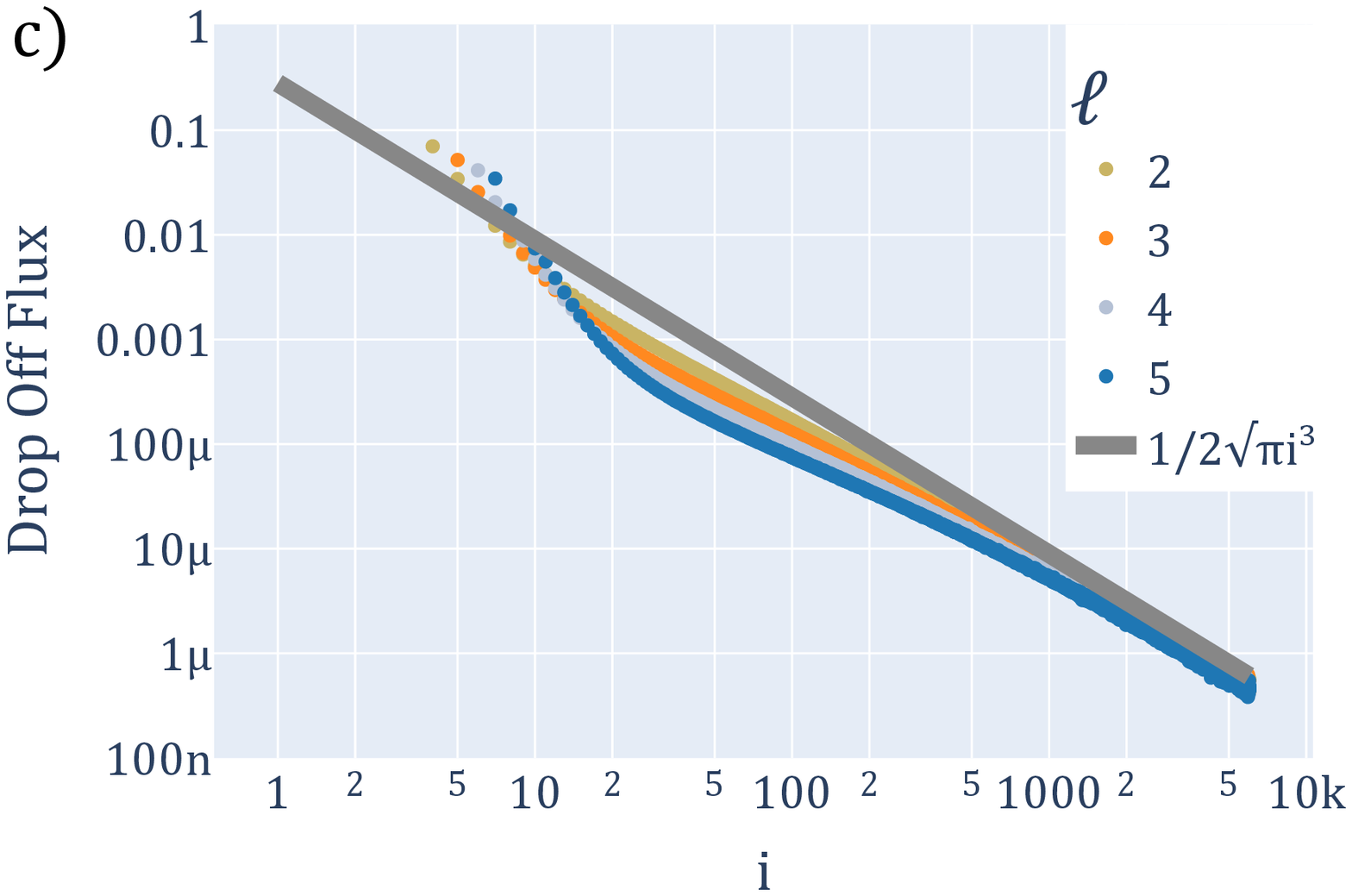}
    \caption{\textbf{Stationary density profile and drop-off flux for extended particles.} 
    The figure shows  simulation results for a homogeneous system of $N=6000$ sites and parameters $\alpha = \beta = \omega = 1$. The first two panels show the stationary density profiles for a) trailing and b) leading particle drop-off for different particle sizes $\ell$. The line shows the prediction for the coalescence model from equation (\ref{eq:density}). Asymptotically the stationary density approaches the profile predicted for $\ell = 1$. c) Stationary drop-off flux for leading particle drop-off and different particle sizes $\ell$. The line shows the prediction 
    $\phi_i \approx \frac{1}{2 \sqrt{\pi i^3}}$ obtained from equations (\ref{eq:density}) and (\ref{eq:dropoffflux}). In both panels the particle size increases from top to bottom. For each particle size eight simulations were performed. Each simulation ran for $10^8$ time steps to reach the steady state followed by $8 \times 10^8$ time steps where the data was collected.
    }
    \label{fig:fig8}
\end{figure}

Figure \ref{fig:fig7} compares simulation results for the collision-induced drop-off model 
with $\ell = 1$ to Eq (\ref{eq:density}). 
As expected, the asymptotic behaviour of the particle density is independent of the initiation rate. Moreover,
the prefactor of the power law decay 
agrees with the prediction for the coalescence model when either the leading or the trailing particle drops off, and with
the prediction for the annihilation model when both particles drop off in a collision. The results in Figs \ref{fig:fig8}a) and b) suggest that, following an $\ell$-dependent crossover region, the \textit{same} asymptotic decay law holds for extended particles with $\ell > 1$. This universal behaviour is indeed expected to appear when the distance between particles is much large than their size. The crossover region is more pronounced for leading particle drop-off, because in that case no drop-off events occur in the initiation region $1 \leq i \leq \ell$. 

The particle density profile (\ref{eq:density}) can be related to the density of nearest-neighbour pairs $\rho^{(2)}_i$ 
using a mass balance argument \cite{Ayyer2010}. Denoting by $n_i \in \{0,1\}$ the occupation number at site $i$, the density $\rho_i = \langle n_i \rangle$ at site $i$ in our model changes in the bulk of the lattice according to 
\begin{equation}
    \label{eq:meanevolution} 
    \fl \qquad \,\,\,\,\, \omega^{-1} \frac{d}{dt} \langle n_i \rangle = \langle n_{i-1} (1-n_i) \rangle - \langle n_i (1-n_{i+1}) \rangle - 
    \delta_\mathrm{lead} \langle n_{i-1} n_i \rangle - \delta_\mathrm{trail} \langle n_{i+1} n_i \rangle. 
\end{equation}
Setting the right hand side equal to zero and using a continuum approximation we arrive at the relation
\begin{equation}
    \label{eq:pairs}
    \rho^{(2)}_i \equiv \langle n_{i-1} n_i \rangle \approx - \frac{1}{\delta_\mathrm{trail} + \delta_\mathrm{lead}} \, \frac{d}{di} \rho_i \sim i^{-\frac{3}{2}}
\end{equation}
to leading order in $i$. Importantly, $\rho^{(2)}_i \ll \rho_i^2 \sim i^{-1}$, which is a consequence of the anomalous kinetics of one-dimensional diffusion limited reactions \cite{Cheng1989} and invalidates the mean-field approach for this class
of systems. 
In the context of the collision-induced drop-off model, $\rho^{(2)}_i$ is proportional to the drop-off flux of particles leaving
the lattice. Specifically, denoting the rate of drop-off from site $i$ by $\phi_i$, we have the relation
\begin{equation}
\label{eq:dropoffflux}
\phi_i = \omega \delta_\mathrm{trail} \langle n_{i} n_{i+1} \rangle +
\omega \delta_\mathrm{lead} \langle n_{i-1} n_i \rangle \approx 
\omega (\delta_\mathrm{trail} + \delta_\mathrm{lead}) \rho^{(2)}_i
\approx - \omega \frac{d}{di} \rho_i
\end{equation}
which also follows directly from the mass balance condition. 
Figure \ref{fig:fig8}c) confirms this prediction for $\alpha = 1$ and different values of the particle size $\ell$.

\section{Conclusion and outlook}

This work was motivated by experiments of 
Park and Subramaniam, who observed a non-monotonic dependence of the rate of protein production on initiation rate for a genetic construct that includes a variable initiation region and a translation bottleneck~\cite{Park2019}. 
We have shown that a simple exclusion model 
is sufficient to explain the main features of the experiment, including in particular the surprising difference between the effects of leading and trailing particle drop-off. An analytically solvable two-particle system reproduces the non-monotonicity of the current in the case of leading particle drop-off, and matches the behaviour of the large-scale system as long as the bulk of the drop-off events occur at the bottleneck. 
The degree of 
non-monotonicity depends on the particle size, because the current reduction per drop-off event is proportional to $\ell$.   
In addition, 
the current displays a complex dependence on the bottleneck length, which reflects the interplay of the mean and variance of the waiting time required for passing the bottleneck \cite{Moffitt2014}.

When drop-off events occur uniformly along the transcript, the ribosome density decreases with the distance from the initiation region. By exploiting the equivalence to previously studied asymmetric reaction-diffusion systems, we have shown that the resulting asymptotic density profile takes the form of an inverse square-root dependence that is universal with respect to the initiation rate and the particle size. This dependence is qualitatively different, and much slower than the exponential decay predicted by models assuming a constant
drop-off rate \cite{Valleriani2010}. The latter prediction has been used for extracting drop-off rates from experimental ribosome profiling data \cite{Sin2016}. It would be of interest to revisit such data to look for signatures of a non-exponential density profile and, thus, of a collective drop-off mechanism.

In the light of the modeling results presented here as well as in \cite{Park2019}, the experimental observation of a non-monotonic dependence
of protein production on initiation rate proves that, at least in the
setting considered by Park and Subramaniam, collision-induced drop-off affects
the leading, rather than the trailing particle of a pair. This appears to make good biological sense: If a ribosome stalls because
it is defective, it will be removed from the transcript by the collision with 
a trailing ribosome. To explore the role of collision-induced 
drop-off as a mechanism of translational quality control, the current investigation could be
extended to include slow ribosomes along the lines of previous studies on 
exclusion models with particle-wise disorder \cite{Krug1996}. Future work should also take
into account the effects of mRNA degradation \cite{Nagar2011}, which can be 
exacerbated by ribosome collisions \cite{Meydan2021,Simms2017,Leedom2022}. 

\section*{Acknowledgements} We thank Juraj Szavits-Nossan and Gunter Sch\"utz 
for useful remarks.

\section*{References}
\bibliographystyle{iopart-num.bst}
\bibliography{references}

\end{document}